\documentclass[useAMS,usenatbib,fleqn]{mn2e}
\usepackage{amsmath,amssymb,dsfont,graphicx,tablefootnote,mathrsfs,float,latexsym,caption,subcaption}
\usepackage{epstopdf,ragged2e}
\usepackage{longtable}
\usepackage{slashed}
\usepackage{fixltx2e}
\usepackage{natbib}

\newcommand{\be}{\begin{equation}}
\newcommand{\ee}{\end{equation}}

\newcommand{\bsym}{\boldsymbol}
\newcommand{\bsb}{\boldsymbol{B}}

\newcommand{\rstar}{R_\ast}
\newcommand{\mstar}{M_*}

\def\04a{{2004 a}}
\def\04b{{2004 b}}

\title{Gravitational radiation from neutron stars deformed by crustal Hall drift}

\author[A.G.~Suvorov, A.~Mastrano , \&  U.~Geppert] {A.G.~Suvorov$^{1}$\thanks{E-mail:suvorova@student.unimelb.edu.au}, A.~Mastrano$^1$, U.~Geppert$^{2,3}$ \\
$^1$ School of Physics, University of Melbourne, Parkville VIC 3010, Australia\\
$^2$ J. Gil Institute of Astronomy, University of Zielona G\'{o}ra, Lubuska 2, 65-265, Zielona G\'{o}ra, Poland \\
$^3$ German Aerospace Center, Institute for Space Systems, Robert-Hooke-Str. 7, 28359 Bremen, Germany}

\begin{document}

\date{}

\maketitle

\label{firstpage}

\begin{abstract}
A precondition for the radio emission of pulsars is the existence of strong, small-scale magnetic field structures (`magnetic spots') in the polar cap region. Their creation can proceed via crustal Hall drift out of two qualitatively and quantitatively different initial magnetic field configurations: a field confined completely to the crust and another which penetrates the whole star. The aim of this study is to explore whether these magnetic structures in the crust can deform the star sufficiently to make it an observable source of gravitational waves. We model the evolution of these field configurations, which can develop, within $\sim 10^4$ -- $10^5$ yr, magnetic spots with local surface field strengths $\sim 10^{14}$ G maintained over $\gtrsim 10^6$ yr. Deformations caused by the magnetic forces are calculated. We show that, under favourable initial conditions, a star undergoing crustal Hall drift can have ellipticity $\epsilon\sim 10^{-6}$, even with sub-magnetar polar field strengths, after $\sim 10^5$ yr. A pulsar rotating at $\sim 10^2$ Hz with such $\epsilon$ is a promising gravitational-wave source candidate.
Since such large deformations can be caused only by a particular magnetic field configuration that penetrates the whole star and whose maximum magnetic energy is concentrated in the outer core region, gravitational wave emission observed from radio pulsars can thus inform us about the internal field structures of young neutron stars.

\end{abstract}

\begin{keywords}
stars: pulsars: general -- stars: neutron - stars: magnetic fields - physical data and processes: gravitational waves
\end{keywords}

\section{Introduction}

Coherent radio pulsar emission is thought to require the creation of a sufficient number of electron-positron pairs in vacuum gaps in the pulsar magnetosphere  \citep{RS75,AS79}. This process takes place in the inner acceleration region just above the polar cap. A precondition for this process is the existence of a small-scale (curvature radius $R_{cur}\lesssim 10^6$ cm) and locally-strong surface magnetic field ($B_s \gtrsim 5\times 10^{13}$ G). The formation of such `magnetic spots' may proceed via crustal Hall drift, which transfers magnetic energy from a strong dipolar toroidal field to small-scale poloidal field structures \citep{GV14}. While the magnetic spots are located just beneath the surface, the toroidal field resides in the deeper regions, in the vicinity of the crust-core interface where the Ohmic diffusion times are comparable to radio pulsar lifetimes. The necessity of such a magnetic spot for radio emission and the process that forms it are described in detail by \citet{GGM13} and \citet{GV14}.

\citet{GV14} identified two very different classes of initial magnetic field configurations that are equally suitable to provide the required magnetic spots within the right time-scale and to maintain them over a radio pulsar lifetime. One of these initial configurations has the field confined within the crust, while the other has the field penetrating the whole star. The maximum toroidal field of the latter is located in the outer core and exceeds that of the former by about one order of magnitude. In both cases, the magnetic energy contribution of the toroidal field component dominates. While the crustal field for both configurations evolves on a time scale of $\sim 10^4$ yr into hemispherically asymmetric structures, the core field remains practically unchanged over radio pulsar lifetimes. This has been recently confirmed by \cite{EPRGV16}, who performed detailed core field evolution studies. In this current paper, two representatives of these different initial field configurations (which return the same polar cap field structure conducive to radio pulsar emission) are taken as an input to explore their effect on the neutron star deformation.

\cite{MSM15} recently presented a method to calculate the deformation of a neutron star caused by poloidal-toroidal magnetic fields consisting of arbitrary multipoles [see also \cite{MLM13}]. In order to explore whether the magnetic spots and strong toroidal fields in radio pulsars produce an ellipticity which is potentially detectable through gravitational wave (hereafter GW) emission, we do not present an exhaustive study of magnetic field structures and their resulting ellipticities; we simply aim to convince the reader that magnetic field structures arising from Hall drift in radio pulsars may induce stellar deformations, which make them potentially detectable as GW sources. In particular, while magnetars have very strong magnetic fields ($B_{\text{pole}} \gtrsim 10^{15}$ G), their spin frequency $\nu\sim 0.1$ Hertz, which results in weak GW luminosities, since the dimensionless GW strain $h_0\propto \nu^2$. Radio pulsars typically have $\nu \sim 100$ Hz, resulting in comparable or larger GW amplitudes if the star is deformed by internal field substructures predicted by the Hall drift simulations.

In Section 2, we discuss the magneto-thermal evolution of two different magnetic field configurations, we recap the method for calculating magnetically-induced density perturbation and, hence, stellar deformation. In Section 3, we discuss the calculated observables for a representative selection of models. In section 4, we summarise and discuss the potential for detecting GW from non-magnetar neutron stars which have undergone Hall drift using current and near-future detectors and compare our results with the current observational upper limits of some known pulsars.

\section{Input and Method}

In this section, we discuss the magnetic field configuration of a neutron star experiencing Hall drift. We present the results of the numerical magneto-thermal simulation by \citet{VPM12} and \citet{GV14} in Section 2.1, we recap the essentials of the density perturbation calculation in a non-barotropic star \citep{MMRA11,MSM15} in Section 2.2, and we discuss the importance of the magnetic dipole moment and how it is calculated in Section 2.3.

\subsection{Magnetic Field Configuration and Evolution}

The magnetic field is evolved using the Alicante group magneto-thermal code \citep{VPM12}. The numerical method and the microphysics are described in detail by \cite{VPM12}. The applicability of this code to model the magneto-thermal evolution in neutron stars for axisymmetric field configurations has been demonstrated by \cite{VRPPAM13}. This code has been applied recently by \cite{GV14} to study the creation of magnetic spots at the neutron star surface for four different assumptions about the stellar magnetic field, in particular classifying the initial topological set-up and respective poloidal and toroidal field strengths. We choose two of these models as representatives: (i) AL, where the magnetic field is dipolar and confined to the crust, and (ii) BL, where the magnetic field lines are dipolar but penetrate the entire star.

While the AL model may be unphysical, it possesses an interesting feature in comparison to the BL model. Both models require strong toroidal fields to generate the magnetic spots, but the AL model forms them with lower initial field strengths relative to the BL model, by $\sim$ one order of magnitude \citep{GV14}. In fact, having a large ($\gtrsim 10^{16} \text{G}$) initial toroidal field strength in the AL model does not result in the formation of a magnetic spot, due to dissipation caused by strong Joule heating. Since there is some evidence that hydrodynamical models with dominant toroidal fields may be unstable [e.g. \cite{aetal13,HK15}], it is worthwhile to consider the AL configuration, which generates magnetic spots with weaker initial toroidal fields.

\begin{figure*}
\centering
\includegraphics[width=12cm]{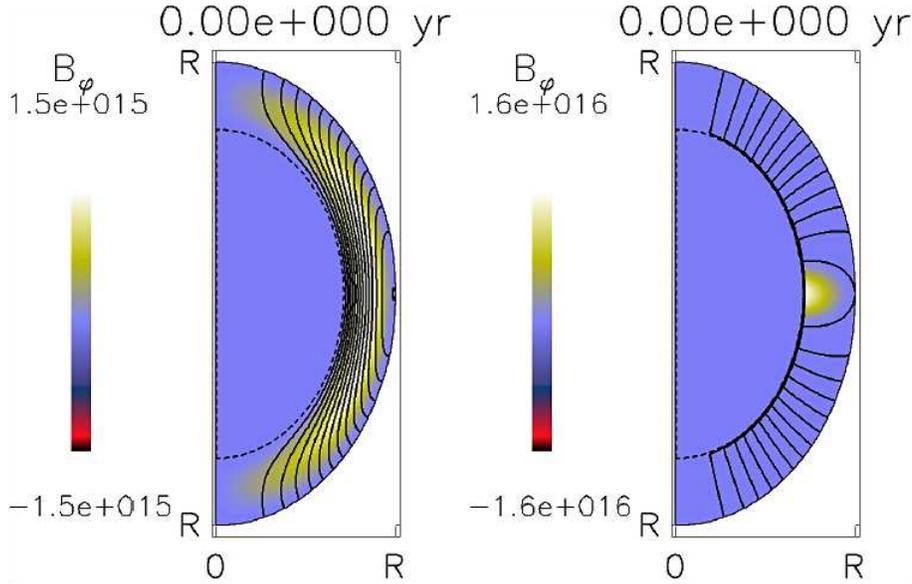}
\caption{Internal field structures for model AL (left) and BL (right) at time $t=0$. For the BL model, only the crustal field lines are shown. The maximum $B_{\varphi}$ values refer also to the crust only; the global toroidal maximum is located in the outer core and remains unchanged for $10^6$ yr. Poloidal field lines are drawn as black solid curves, and colours map the intensity of the toroidal magnetic field $B_{\varphi}$. For better visibility, the crust is stretched by a factor of $4$ in the image.}
\label{fig:Initial_B_global}
\end{figure*}
	
\begin{figure*}
\centering
\includegraphics[width=12cm]{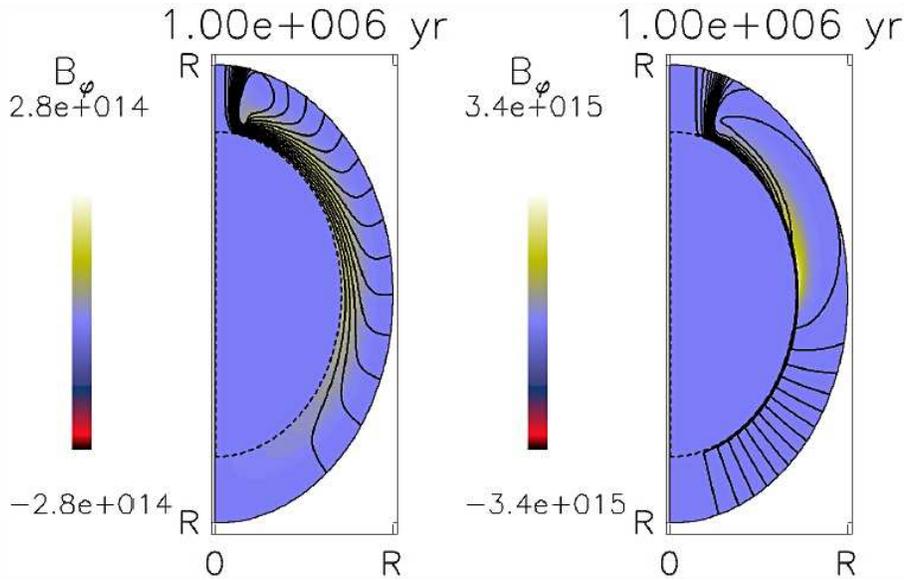}
\caption{Internal field structures for model AL (left) and BL (right) at time $t=10^6$ yr. For the BL model, only the crustal field lines are shown. The maximum $B_{\varphi}$ values refer also to the crust only; the global toroidal maximum is located in the outer core and remains almost unchanged for $10^6$ yr. Poloidal field lines are drawn as black solid curves, and colours map the intensity of the toroidal magnetic field $B_{\varphi}$. For better visibility, the crust is stretched by a factor of $4$ in the image.}
\label{fig:B_global_1e6}
\end{figure*}

The magnetic 4-field components are given by

\begin{equation} \label{eq:magfield22}
B^{\mu} = \frac {1} {2} \epsilon^{\mu \nu \kappa \lambda} u_{\nu} F_{\lambda \kappa},
\end{equation}
where $\boldsymbol{F}$ is the Faraday tensor, and $\boldsymbol{u}$ is the 4-velocity \citep{L67}. In the crust, the magnetic 3-field $\bsb$ is tied to the electrons which circulate in currents through a crystalline lattice, formed by almost immobile ions. Therefore, the only processes that drive the magnetic evolution are Ohmic diffusion/dissipation and Hall drift, described by the Hall induction equation (see e.g. \citealt{GR92,PG07}):

\begin{equation}
\frac{\partial\boldsymbol {B}}{\partial t}= - \slashed{\nabla} \times\left[\frac{c^2}{4\pi\sigma}\slashed{\nabla} \times (e^{\nu}\boldsymbol{B}) +
\frac{c}{4\pi e n_e} [({\slashed{\nabla}} \times (e^{\nu}\boldsymbol{B})] \times \boldsymbol{B}  \right],
\label{eq:Hallind}
\end{equation}
where $\sigma$ denotes the electric conductivity, which depends on the local temperature $T$, the density $\rho$, and the composition of the crustal ionic lattice. The electrical 3-current is given by $\boldsymbol{J}=ce^{-\nu}{\slashed{\nabla}}\times (e^{\nu}\bsym{B})/4\pi$. The second term in the right-hand side of equation~(\ref{eq:Hallind}) represents the Hall drift, whose pre-factor depends on the crustal electron number density $n_e$. The $\slashed{\nabla}$-operator represents the spatial $3$-covariant derivative, taken with respect to the usual spherically symmetric Oppenheimer-Volkoff metric where the curvature of space is taken into account through the mass distribution of the star [see e.g. \cite{W84}]. The gravitational redshift factor $e^{\nu}$ forms the $tt$-component of the metric and is given by the structure of the star as a solution to the Einstein equations \citep{GPZ00}. Thus, general relativistic effects influence the magnetic field evolution in three ways, namely via the presence of the red shift factor, its spatial derivative, and the intrinsically curved nature of the space, modifying the $\slashed{\nabla}$-operator. All coefficients in equation~(\ref{eq:Hallind}) are functions of the radial coordinate $r$ and time $t$.

The mutuality of thermal and magnetic evolution is seen in the energy equation that
describes the evolution of the crustal temperature $T$, viz.

\begin{equation}
c_v e^{\nu}\frac{\partial T}{\partial t} - {\slashed{\nabla}} \cdot \left[e^{\nu}\hat{\kappa}\cdot {\slashed{\nabla}}(e^{\nu}T)\right]
=e^{2\nu}\left(-Q_{\nu}+Q_h\right)~,
\label{eq:Tevol}
\end{equation}
where $c_v$ is the specific heat, $Q_\nu$ is the neutrino luminosity, and $Q_h=|\boldsymbol{J}|^2/\sigma$ is the Joule heating. Equation \eqref{eq:Tevol} is strongly coupled to the magnetic evolution through the $\bsb$-dependent components of the heat conductivity tensor $\hat\kappa$ \citep{GKP04}, $Q_h$ \citep{PMG09,VRPPAM13}, and, to a lesser extent, by the weak $\bsb$-dependence of the processes contributing to $Q_{\nu}$.

For the AL model, we assume that the crustal field cannot penetrate into the core. The magnetic field decays smoothly to zero at the crust-core boundary. The density and pressure are glued to their initial values at this interface. The boundary condition in the BL model, in contrast to the AL model, is that the magnetic field is continuous across the crust-core interface.

For the BL model, the evolution of the \emph{core} magnetic field is certainly not correctly described  by equation \eqref{eq:Hallind}. Ambipolar diffusion, as well as processes occurring in superfluid/superconducting matter play a more important role in the evolution of the core field than Ohmic diffusion and Hall drift \citep{HRV08,GAGL15}. However, several numerical investigations suggest that the core field does not evolve significantly on timescales of the order $10^{6}$ yr (see \citep{EPRGV16} and references therein). Therefore, we take the simplifying assumption that the core magnetic field remains unperturbed during the Hall-drift-driven crustal field evolution. In this paper, we wish to focus on the formation of the magnetic spots in the crust and their role in deforming the star. This is justified also because hemispherically asymmetric crustal magnetic field can create significant ellipticities; they are more spread-out and, in comparison to core matter, less dense crustal material is more susceptible to magnetic forces.

Figures 1 and 2 show the topological structure of the magnetic fields for the AL (left panel) and BL (right panel) model at times $t=0$ and $t=10^{6}$ yr, respectively. The core remains, at all times, unmagnetized for the AL model. For the BL model, the bulk of the magnetic energy is stored within the outer core region, and we assume that the core field structure does \textit{not} change on the time-scales involved here. Therefore, deformations can be induced only by changes of the crustal field configuration which is presented in these figures. Note that since the core magnetic field in the BL model remains unchanged over the simulation time-scale, we opt not to draw the core field lines of the BL configuration in Figs. 1 and 2, for the sake of clarity and focus. In both AL and BL cases, the evolution clearly demonstrates a qualitative similarity in the formation of a spot in the northern hemisphere, as seen by the collection of poloidal field lines. In the southern hemisphere, the AL model still has a strong toroidal component that lingers on after $10^{6}$ years, while the toroidal component of the BL model is entirely concentrated in the northern hemisphere. Note that the inclusion of the negative values of $\boldsymbol{B}_{\phi}$ in the colour scale is a plotting artefact associated with the Alicante code. For the initial magnetic field configurations considered here (both the poloidal and the toroidal field component are dipolar), $\boldsymbol{B}_{\phi}$ is predominantly positive throughout the star.

At first glance, it may seem inappropriate to compare the AL and BL models to each other, since their initial configurations are clearly different and their initial field strengths are $\sim$ one order of magnitude apart. The only criterion we applied when choosing them as representatives of the respective classes of initial field configurations is their ability of generating magnetic spots of sufficient strength ($\sim 10^{14}$ G) within $\sim 10^4$ yr and maintaining them up to $\sim 10^6$ yr. We have therefore chosen to focus on these particular configurations and initial field strengths and compare their Hall-drift-driven evolutions in this paper.

\subsection{Field Induced Neutron Star Deformations}

In this section we detail the calculation of the stellar ellipticity for a given \textit{analytic} magnetic field. We find that using the raw output from the Alicante code to calculate the stellar deformation results in an unacceptable amount of error, since this computation involves taking high order derivatives and integrals of the components of the magnetic field (see \eqref{ellipticity} below). Therefore, we take a brief detour here to discuss how one calculates the deformation given an analytic field. In section 3.2, we show how one reconstructs an analytic representation of any given numerical magnetic field output on a grid. In particular, we build an analytic replica of the Alicante output, whose derivatives and integrals are then computed without introducing additional errors, such as those that would come from using, e.g., Simpson's rule.

Given the spatial components of $\bsb$ from \eqref{eq:magfield22}, we can calculate the ellipticity. We begin by decomposing the magnetic field into its poloidal and toroidal components and express it in dimensionless spherical polar coordinates ($r,\theta,\phi$), such that the stellar surface is located at $r=1$ \citep{c56,MMRA11,MM12,MLM13}, viz.

\be {\bf{B}}=B_0 [\eta_p \nabla\alpha(r,\theta)\times \nabla\phi + \eta_t \beta(\alpha)\nabla\phi],\label{eq:magf} \ee
where $B_0$ parametrizes the overall strength of the field, $\eta_{p}$ and $\eta_t$ set the relative strengths of the poloidal and toroidal components respectively ($\eta_p= 1$ without loss of generality), $\alpha(r,\theta)$ is the poloidal magnetic stream function, and the function $\beta(\alpha)$ defines the toroidal field component. Note that we use the $\nabla$-operator here to represent the usual 3-dimensional Euclidean gradient operator. We require the analytic field (i) to be symmetric about the $z$-axis, (ii) to be current-free and purely poloidal outside the star, (iii) to have a poloidal component that is continuous everywhere, and (iv) to yield finite current everywhere (which vanishes at the stellar surface). These conditions are to be fulfilled by judicious choices of $\alpha$ and $\beta$.

The magnetic energy density is $\lesssim 10^{-6}$ of the gravitational energy density, even in magnetars. Therefore, we can treat the magnetic force as a perturbation\footnote{Note that we model the Eulerian density and pressure perturbations as Newtonian quantities, while the magnetic field evolution is governed by a relativistic induction equation \eqref{eq:Hallind}. While these are in principle incompatible assumptions, we find that treating the perturbations in $\rho$ and $p$ arising from $\boldsymbol{B}$ as a Newtonian system introduces negligible errors in the calculation of observables (see Appendix A).} on a background hydrostatic equilibrium and write the Newtonian hydromagnetic force balance equation as

\be\frac{1}{4\pi} (\nabla\times {\bf{B}})\times{\bf{B}}=\nabla\delta p +\delta\rho\nabla\Phi,\label{eq:forcebal}\ee
to first order in $B^2/(\mu_0 p)$ in the Cowling approximation $(\delta\Phi = 0)$, where $p_0$ is the zeroth-order pressure, $\rho_0$ is the zeroth-order density, $\Phi$ is the gravitational potential, and $\delta p$, $\delta\rho$, $\delta\Phi$ are perturbations of the latter three quantities. Because we do not assume a barotropic star,\footnote{For a discussion on the applicability of the non-barotropic assumption to neutron stars, see Sec. 2 of \cite{MSM15}.} the density perturbation $\delta\rho$ does not have to be a function solely of the pressure perturbation $\delta p$, and therefore the equation of state imposes no restrictions on the field structure. Physically, this means that the imposed magnetic field sets the density and pressure perturbations, but the resulting perturbations do not restrict the magnetic field in turn. Therefore, we do not specify a barotropic equation of state and then solve the Grad-Shafranov equation for the magnetic field configuration. Instead, we specify the magnetic field whose effects we wish to investigate, then calculate the density perturbations that the field causes. The method used here to specify density and pressure is unphysical in its simplicity, though is still, to leading order, an accurately representation of the deformation induced by the magnetic field \citep{MSM15}.

We characterize the magnetic deformation of the star by its ellipticity $\epsilon$,

\be \label{eq:epsilon} \epsilon = \frac{I_{zz}-I_{xx}}{I_0},\ee
where $I_0$ is the moment of inertia of the unperturbed spherical star, the moment-of-inertia tensor is given by

\be \label{eq:deltarho} I_{jk} = R^5_* \int_V \textrm{d}^3x[\rho(r)+\delta\rho(r,\theta)](r^2\delta_{jk}-x_j x_k),\ee
$R_*$ is the stellar radius, and the integral is taken over the volume of the star $(r\leqslant 1)$. The density perturbation $\delta\rho$ is calculated by taking the curl of both sides of equation (\ref{eq:forcebal}) and matching the $\phi$-components:

\be \frac{\partial\delta\rho}{\partial\theta}=-\frac{r}{4\pi \rstar}\frac{\mathrm{d}r}{\mathrm{d}\Phi}\{\nabla\times[(\nabla\times{\bf{B}})\times{\bf{B}}]\}_\phi.\label{ellipticity}\ee
Equations (\ref{eq:epsilon})--(\ref{ellipticity}) are then solved to obtain $\epsilon$.

\subsection{Magnetic Dipole Moment}

While we demonstrate in this paper that neutron stars undergoing Hall drift may produce significant $\epsilon$, it is useful to consider observational counterparts in the electromagnetic spectrum also. A neutron star undergoing Hall drift experiences a lowering of its total magnetic energy (see Sec. 3.5). Assuming this energy to be associated with electromagnetic braking torque (and gravitational radiation), one can place bounds on the magnetic dipole moment $|\mu|$ from spin down measurements [e.g., \citep{MEL97}].

The boundary conditions for the magnetic field (Sec. 2.2) ensure that the surface current vanishes. The magnetic field at the stellar surface is therefore uniquely determined by its radial component $B_{r}$ \citep{BC54}, which satisfies

\begin{equation}
\nabla^{2} B_{r} = 0,
\end{equation}
by Maxwell's equations. After performing a multipole expansion and extracting only the dipole moment, we find \citep{VM08},

\begin{equation}
|\boldsymbol{\mu}| = \frac {3 R_{\star}^{3}} {4} \int^{1}_{-1} d (\cos \theta) \cos \theta B_{r}(1,\theta).
\end{equation}

The emergence of high-order multipoles, induced by the Hall drift, means that the dipole moment can change as the star evolves, independent of how $\epsilon$ evolves.

\section{Calculation of observables}

In this section, we analyze the field structures of the AL and BL models after undergoing Hall drift towards equilibrium to estimate the magnitude of associated gravitational radiation. We choose a range of initial poloidal and toroidal field strengths, show how the magnetic field evolves, model the field analytically, and use the analytic model to calculate $\delta\rho$, $\epsilon$, and $\mu$.


\subsection{Hall evolution of the magnetic field}

After typically $\sim 10^5 $ yr, the crustal magnetic field settles into a Hall equilibrium which is characterized by the presence of a strong, localized magnetic spot in the northern hemisphere. The strength and location of the magnetic spot depend on the initial field structure. As an example, we simulate numerically the magneto-thermal evolution of an AL and a BL model star. Both the poloidal and the toroidal components are initially dipolar, as shown in Fig. \ref{fig:Initial_B_global}, on a 100 radial $(r)$ points by 180 angular $(\theta)$ points grid. We list the crustal maximum poloidal field strength $|\bsb_\textrm{pol}^{\textrm{max}}|$, the maximum toroidal field strength $|\bsb_\textrm{tor}^{\textrm{max}}|$, and the locations of these maxima at $t=0$, $10^3$, $10^4$, $10^5$, and $10^6$ yr in the first five columns of Tables \ref{tab:table1} (for the AL model) and \ref{tab:table2} (for the BL model). The final field configurations at $t=10^6$ yr are shown in  Fig. \ref{fig:B_global_1e6}. We gloss over the details of the simulations themselves, since the focus of this paper is on the analytic modelling of $\bsb(r,\theta)$, the calculations of $\delta\rho$, $\epsilon$, $\mu$, and their evolution over time. We refer the reader to \citep{GV14} and references therein for more details about the simulations.

In both models, the magnetic spot develops within $\sim 10^5$ yr. However, in the BL model, the magnetic spot is located closer to the equator than in the AL model, and the surface field strength at the spot is $\sim 5$ times higher. This is a consequence of the initial toroidal field, which is both stronger and deeper in the star than in the AL model. The poloidal field taps into the magnetic energy of the toroidal field via Hall drift. Since the initial toroidal field is stronger in the BL model than in the AL model, the resulting magnetic spot of the BL model is stronger.

We reconstruct these fields analytically in Sec. 3.2, calculate $\epsilon$ and $\mu$ in Sec. 3.4 below, and list the results in the last two columns of Tables \ref{tab:table1} and \ref{tab:table2}.

\subsection{Analytic field reconstruction}
As discussed in section 2.2, the calculation of the density perturbation due to the magnetic field in the non-barotropic approach, as in equation \eqref{eq:deltarho}, involves a number of differentiations of terms which are nonlinear in the components of $\boldsymbol{B}$, followed by a subsequent symbolic integration with respect to $\theta$. In an effort to minimise errors obtained through these differential operations, we reconstruct analytically the magnetic field given by the Alicante code to an accuracy within a few percent (see Sec. 3.3).

We begin by expanding the stream function $\alpha$ in equation \eqref{eq:magf} in the usual multipole series up to order $N$ weighted by radial functions,

\begin{equation} \label{eq:multmom}
\alpha(r,\theta) = \sum_{\ell = 1}^{N}  \kappa_{\ell} f_{\ell}(r) Y_{\ell 0}'(\theta) \sin \theta ,
\end{equation}
where $Y_{\ell}$ are the spherical harmonics and the functions $f_{\ell}$ are subject to the boundary conditions presented in Sec. 2.2 [see Sec. 4.1 of \cite{MLM13}]. The task now is to obtain an algorithm for finding constants $\kappa_{\ell}$ and functions $f_{\ell}$ such that the field \eqref{eq:magf} matches the output of the Alicante code within a specified tolerance at each grid point.



For simplicity and to ensure that the boundary conditions (Sec. 2.2) can be satisfied for each $\ell$, the functions $f_{\ell}$ are taken to be polynomials with even powers $4,6,\cdots,m$, i.e.,

\begin{equation}
f_{\ell}(r) = \sum_{i=4,6,8,\cdots}^{m} a_{i \ell} r^{i}.
\end{equation}
Enforcing the boundary conditions amounts to adding some algebraic constants on the $a_{i \ell}$. These constraints determine three such $a_{i \ell}$ for each $\ell$ [boundary conditions (ii)--(iv) in Sec. 2.2], leaving $(1/2)(m-8)$ undetermined. Finally, we choose the toroidal function\footnote{This choice results in the toroidal field being defined nominally everywhere, which is an unphysical assumption. However, we may place additional constraints on the constants $a_{i \ell}$ such that $\boldsymbol{B}_{\phi}$ is small everywhere except in regions of interests, thereby not influencing calculated observables. These constraints do not conflict with our need for matching the poloidal components since the poloidal and toroidal components are linearly independent. Incidentally, such a choice of $\beta$ removes difficulties associated with ensuring that the perturbed density profile $\delta \rho$ is continuous around the neutral curves of the multipolar poloidal field [see \cite{MLM13,MSM15}].} $\beta$ in equation \eqref{eq:magf} to be of power-law form, $\beta = \alpha^{\gamma}$, for some real number $\gamma$. We therefore have $N$ free parameters from $\kappa_{\ell}$, $(1/2) (m-8)$ free parameters from $f_{\ell}$, two  parameters from matching $B_{0}$ and $\eta_t$, as well as $\gamma$, making a total of $[(N/2)(m-8)]+ 3 $ to work with in matching \eqref{eq:magf} with the numerical Alicante grid.

To choose these constants, we perform a least squares analysis. We build a numerical 3-index object $M^{N}_{ijk}$ which contains all information pertaining to the components of the Alicante magnetic field $\boldsymbol{B}^{N}_{k}$ at each grid point $(i,j)$. Explicitly, the components of $\boldsymbol{M}^{N}$ are given by
\begin{equation} \label{eq:matrixn}
M_{ijk}^{N} = \boldsymbol{B}^{N}_{k}(r_{i},\theta_{j}),
\end{equation}
where $(r_{i},\theta_{j})$ are the coordinate values of the enumerated grid point $(i,j)$. The analytic field, constructed from \eqref{eq:multmom}, is then evaluated at each grid point $(i,j)$ to produce another 3-index object $M^{A}_{ijk}$ which contains the appropriate arrangement of the $[(N/2)(m-8)]+ 3$ parameters, i.e.,
\begin{equation} \label{eq:matrixa}
M_{ijk}^{A} = \boldsymbol{B}^{A}_{k}(r_{i},\theta_{j}),
\end{equation}
where $\boldsymbol{B}^{A}$ is given by \eqref{eq:magf} together with \eqref{eq:multmom}. We minimize the least squares residuals

\begin{equation}
s_{ijk}^2 = |M^{N}_{ijk} - M^{A}_{ijk}|^{2},
\end{equation}
in the standard manner \citep{bds69}.

In practice, one chooses $N$ and $m$ by an iterative procedure. We find typically that $m = 20$ and $N = 65$ results in maximum residuals that are at most $2 \%$ of $|M^{N}_{ijk}|$, for both the AL and BL models. As an example, for a $100 \times 180$ grid, we have a total of $393$ free parameters with $m=20$ and $N=65$, yielding a ratio of $45.08$:$1$ of grid points to free parameters.

Note that the method presented in this section is not limited to the Alicante grids. This least squares fitting method should be applicable to the analytic reconstruction of any numerically defined magnetic field.

\subsection{Error analysis}

To estimate the maximum errors introduced through the fitting procedure described in the previous section, we determine the value of the relative error $\boldsymbol{\delta}$ defined as

\begin{equation}
\delta_{k} = \max_{i,j} \left( \frac {s_{ijk}} {M^{N}_{ijk}} \right),
\end{equation}
for each $k$. The error is then determined by performing the integration in (7) with respect to the modified components $\boldsymbol{B}_{k} \rightarrow \boldsymbol{B}_{k} \delta_{k}$. For higher resolution $100 \times 180$ Alicante grids, we find that $\delta_{r} = 0.03$, $\delta_{\theta} = 0.03$ and $\delta_{\phi} = 0.02$. The errors in the ellipticity in equation \eqref{eq:epsilon}, which are weighted by the relative strengths of the poloidal and toroidal components (i.e. by the value of $\eta_{t}$), are found to be at most $6 \%$. For`lower' resolution\footnote{Note that while the resolution is lowered, the convergence of the Alicante code is guaranteed due to the nature of the staggered grid arrangement, as detailed in Sec. 3 of \citep{VPM12}. In particular, use of smaller grids requires one to reduce the size of the time step to ensure that the Courant condition is satisfied. The use of lower resolution grids introduces a negligible error into the convergence of the numerical $\boldsymbol{B}$ field, but results in less points being available for the multipole fitter.} $50 \times 90$ Alicante grids, the situation is only slightly worse since we find $\delta_{k} \sim 0.04$ for each $k$, resulting in a maximum error in the ellipticity of $8 \%$. Given the physical uncertainties in the equation of state (or first-order non-barotropic approach), an error of $8 \%$ in the ellipticity appears acceptable.

\subsection{Evolution of $\epsilon$ and $\mu$}

Following the procedure in sections $2$ and $3.2$, we reconstruct the field configuration analytically and calculate $\delta\rho$, $\epsilon$, and $\mu$. The zeroth-order density profile $\rho(r)$ is chosen to be that of an $n=1$ polytropic star [unlike, e.g., \citet{MMRA11} and \citet{MSM15}]

\be  \label{eq:polytrope} \rho = \frac{\rho_\mathrm{c}\sin(\pi r)}{r},\ee
\be p = k\rho^2,\ee
where $\rho_\mathrm{c} = \mstar/(4\rstar^3)$ and $k = (2 G\rstar^2)/\pi$. Using this profile minimises errors from approximating the curved-space magnetic field \eqref{eq:magfield22} with the flat-space one \eqref{eq:magf}, see Appendix A.

In Fig. 3 we present a contour plot of the analytically reconstructed $\delta\rho$ profile for the AL model after $10^{6}$ yr. We list the values of $\epsilon$ and $\mu$ at each time step in the last two columns on Table \ref{tab:table1}. We see that large $(\delta \rho / \rho \sim 10^{-4})$ density perturbations are spread out through the entire crust, with little suggestion that the toroidal field is dominating. The magnetic spot which develops near the north pole deforms the star into an oblate shape $(\epsilon > 0)$ (see Table \ref{tab:table1}).
Despite the presence of the magnetic spot, the density perturbation is seen to be almost hemispherically symmetric. This suggests that the presence of the toroidal field in the southern hemisphere mitigates the effects of the deformation induced by the magnetic spot (compare Figures 2 (left) and 3). For model AL, we find that the dipole moment tends to decrease between $t=0$ and $t=10^{6}$ yr. The ellipticity similarly decreases uniformly. The crustal maximum of the toroidal fields decreases significantly over time (from $\sim 10^{15}$ G to $\sim 10^{14}$ G), indicating that the diffusion of toroidal energy is a universal effect of the Hall drift. The reduction of the dipole moment (from $\sim 8 \times 10^{30}$ G $\text{cm}^3$ to $5 \times 10^{29}$ G $\text{cm}^3$) indicates that magnetic energy is being transferred from the dipole component to higher order multipoles (see Sec. 3.5).

In contrast, model BL shows the opposite development. Figure 4 illustrates a contour plot of the analytically reconstructed $\delta\rho$ profile for the BL model after $10^{6}$ yr. We list the values of $\epsilon$ and $\mu$ at each time step in the last two columns on Table \ref{tab:table2}. The toroidal field is pronounced and penetrates the crust, leading to a greater deformation, which increases uniformly until $10^{5}$ years have elapsed, at which point it stabilises. The Hall drift does not extend into the core, which means that the strong toroidal field present there ($\sim 10^{16}$ G) remains untapped. The deformation induced by the changing magnetic field topology in the crust is then amplified by the core toroidal field which adjusts at the crust-core interface to ensure continuity. The crustal toroidal field is more  pronounced in the northern hemisphere than the southern hemisphere [compare Figures 2 (right-hand panel) and 4], $\delta\rho/\rho$ in the northern hemisphere is $\sim 10^3$ times that of the southern hemisphere. As a result the deformation induced by the magnetic spot is bolstered by the toroidal field, as opposed to diminished in the AL case. The field now deforms the star into a prolate shape $(\epsilon < 0)$. Over time, the star becomes more prolate as the toroidal field develops a maximum around the equator, in spite of the increasing poloidal field strength, which tends to deform the star into an oblate shape (Table \ref{tab:table2}). The dipole moment also increases uniformly. 

There are two major differences between the AL and BL models described in Tables \ref{tab:table1} and \ref{tab:table2}: the presence of a core-penetrating field (BL), and the strength of the magnetic field at $t=0$. We cannot conclude which of the two has a more pronounced effect on the overall deformation without examining each difference in isolation. To explore how the evolution depends on the initial state of the AL and BL models, we perform a series of `lower'-resolution runs of the Alicante magneto-thermal code (50 radial points by 90 angular points) for some combinations of initial poloidal and toroidal field strengths [listed in columns 3 and 5 of Tables \ref{tab:table3} (AL) and \ref{tab:table4} (BL)], which still ensure the presence of the magnetic spot. We show the relevant parameters involved for the initial and final ($t=10^6$ yr) states for the AL model and the BL model in Tables \ref{tab:table3} and \ref{tab:table4}, respectively, in the same format as Tables \ref{tab:table1} and \ref{tab:table2}. For model AL, we see that $\mu$ decreases dramatically, by two orders of magnitude over $10^6$ yr, while the star becomes less prolate. For model BL, we see that $\mu$ does not change much over $10^6$ yr, while the star becomes more prolate. Note also that, for model AL, both $|\boldsymbol{B}^{\text{max}}_{\text{pol}}|$ and $|\boldsymbol{B}^{\text{max}}_{\text{tor}}|$ tend to decrease with time, except for AL (III) and AL (IV), where the initial poloidal maxima are $5\times 10^{13}$ G. On the other hand, for model BL, $|\boldsymbol{B}^{\text{max}}_{\text{tor}}|$ decreases by $\approx 80\%$ while $|\boldsymbol{B}^{\text{max}}_{\text{pol}}|$ increases, even by two orders of magnitude in model BL (III). As in the case of the higher resolution runs, the Hall drift tends to redistribute the magnetic field evenly amongst the poloidal and toroidal components; the larger the initial $|\boldsymbol{B}^{\text{max}}_{\text{tor}}|$, the larger the final $|\boldsymbol{B}^{\text{max}}_{\text{pol}}|$.

Our results thus indicate that crustal Hall drift is most effective at deforming the star when the magnetic field penetrates the entire star (i.e., the BL model). Note that one would indeed expect a stronger magnetic field to lead to a greater deformation since $\epsilon \sim B^2$. As a result, since the BL model begins with an order of magnitude stronger toroidal field, this conclusion is somewhat obvious. Quantitatively, however, at $t=10^{6} \text{yr}$ we see that the ellipticity can be four orders of magnitude weaker for the AL case than the BL case, even though the magnetic field is only one order greater. As a result, the difference in topological structure allows for two orders of magnitude discrepancy, and so the conclusion is truly based on the presence of a core-penetrating field as opposed to simply a stronger initial set-up. This conclusion is also supported by the set of secondary runs, as detailed in Tables \ref{tab:table3} and \ref{tab:table4}, where the BL ellipticity for the lowest initial field strengths [BL (IV)], after $t = 10^{6}$ yr, is greater than the AL ellipticity for the largest initial field strengths [AL (I)].

\begin{figure*}
\centering
\includegraphics[width=14cm]{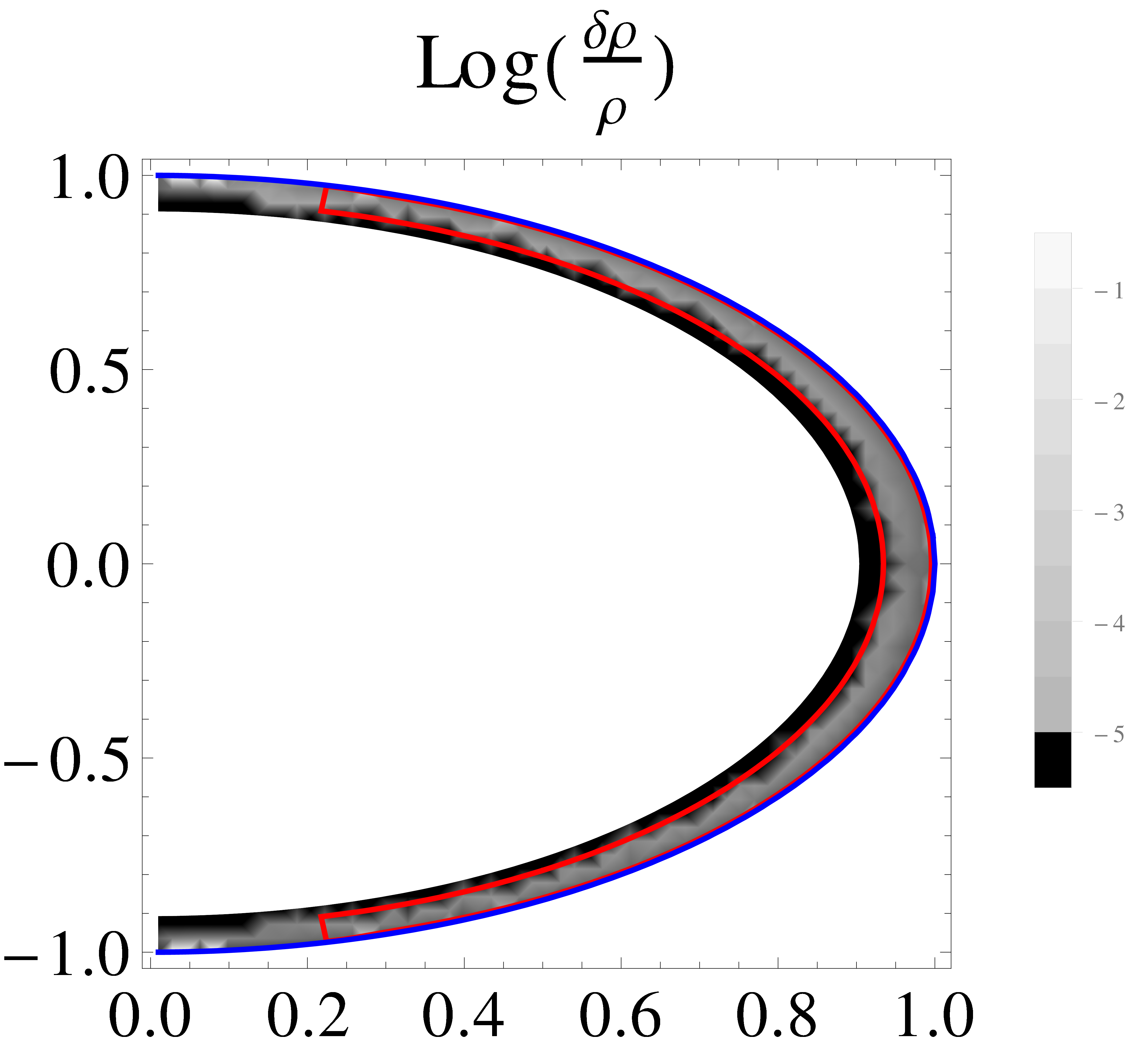}
\caption{Density perturbation for the AL model at $t=10^{6}$ yr corresponding to $|\boldsymbol{B}^{\text{max}}_{\text{pol}}| = 10^{13} \text{G}$ and $|\boldsymbol{B}^{\text{max}}_{\text{tor}}| = 1.5 \times 10^{15} \text{G}$ initially. The stellar surface is indicated by the blue curve, and the toroidal region is enclosed by the red surface. Darker shades indicate a weaker deformation. Note that the core is not shown since the deformation there is zero.}
\label{fig:A_rho_global_1e6}
\end{figure*}

\begin{figure*}
\centering
\includegraphics[width=14cm]{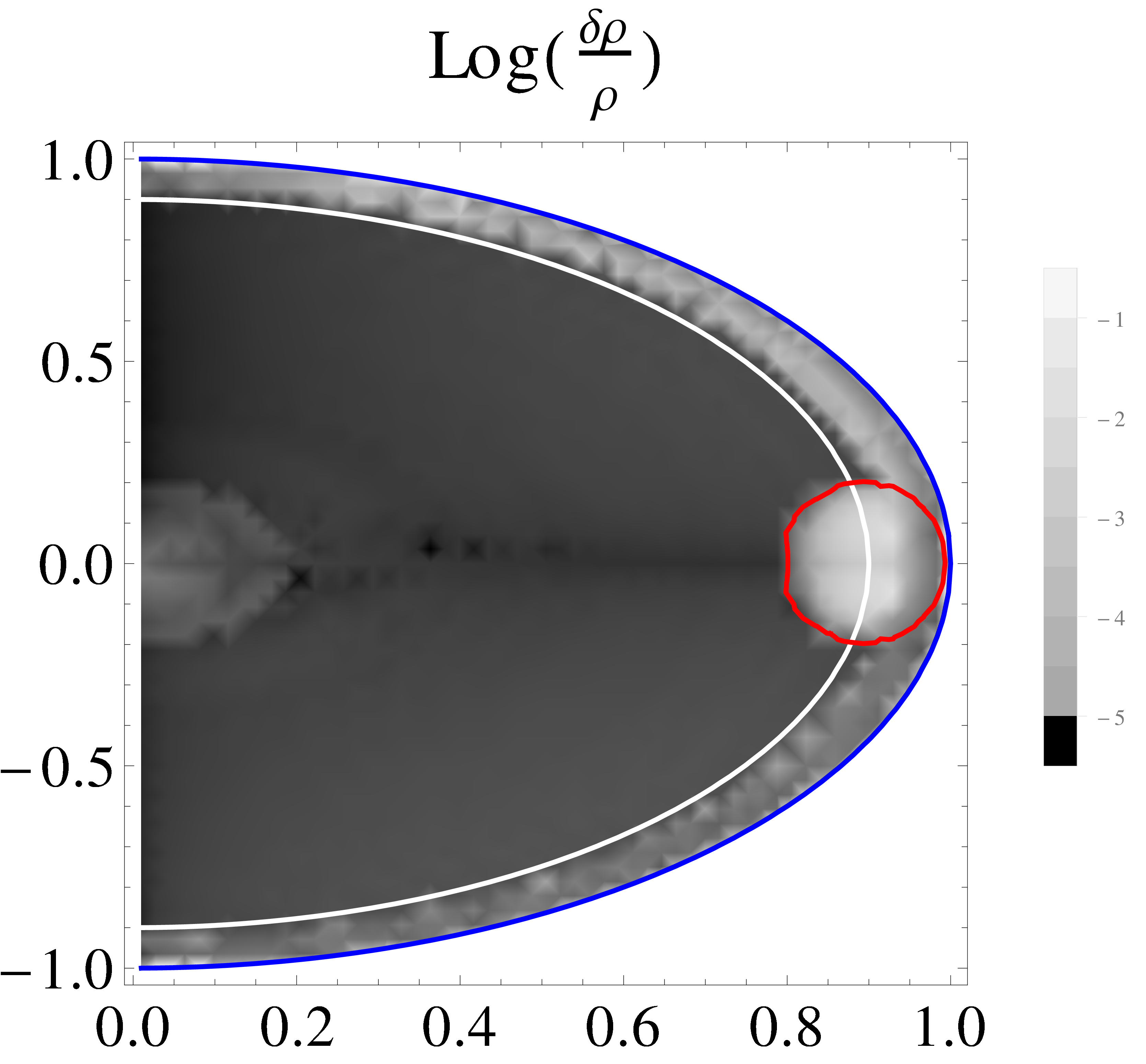}
\caption{Density perturbation for the BL model at $t=10^{6}$ yr corresponding to $|\boldsymbol{B}^{\text{max}}_{\text{pol}}| = 10^{13} \text{G}$ and $|\boldsymbol{B}^{\text{max}}_{\text{tor}}| = 2 \times 10^{16} \text{G}$ initially.  The stellar surface is indicated by the blue curve, the crustal region by the white curve, and the toroidal region is enclosed by the red surface, which penetrates the crust. Darker shades indicate a weaker deformation.}
\label{fig:B_rho_global_1e6}
\end{figure*}

\begin{table*}
\centering

  \caption{Summary of the properties of a particular model AL (where the magnetic field is confined to the crust) at different stages of evolution. We show the maximum crustal field strengths and their respective locations, as well as the ellipticity and magnetic dipole moment. The locations of the maxima are given in dimensionless spherical coordinates $(r,\theta)$, such that $r=1$ is the stellar surface. These runs were obtained for a star with mass $1.4M_{\odot}$. The stellar radius is taken to be $11.6$ km, the crust core interface is located at $10.8$ km, and the bottom of the envelope, up to which equation \eqref{eq:Hallind} has been integrated, is at $11.53$ km. The dimensionless coordinate $r$ is normalised so that at a radius of $11.53$ km we have $r=1$. Therefore the crust-core boundary is located at $r= 0.94$.}
  \begin{tabular}{lcccccc}
  \hline
Time & $|\boldsymbol{B}^{\text{max}}_{\text{pol}}|$  & Crustal location  & $|\boldsymbol{B}^{\text{max}}_{\text{tor}}|$  & Crustal location  & Ellipticity & Dipole moment \\
(yr) & ($10^{14}$ G) & & ($10^{15}$ G) & & & ($10^{30}$ G cm$^3$)\\
\hline
$0$ & $  1.1$  & $(0.94,1.58)$ & $1.5$  & $(0.97,1.58)$ & $-1.09 \times 10^{-8}$ & $7.66 $  \\
$10^{3}$   & $ 1.1 $ & $(0.94,1.58)$ & $ 1.5 $ & $(0.97,1.49)$ & $-5.83 \times 10^{-8}$ & $7.20 $  \\
$10^{4}$   & $ 3.4 $ & $(0.97,0.4)$ & $ 1.4 $ & $(0.96,1.07)$ & $9.57 \times 10^{-10}$ & $5.00 $  \\
$10^{5}$  & $ 17 $ & $(0.94,0.14)$ & $ 1.1 $ & $(0.94,0.47)$ & $1.29 \times 10^{-8}$ & $2.81 $  \\
$10^{6}$  & $ 2.7 $ & $(0.94,0.19)$ & $ 0.28 $ & $(0.94,0.30)$ & $9.93 \times 10^{-10}$ & $0.48 $  \\

\hline
\end{tabular}
\label{tab:table1}
\end{table*}

\begin{table*}
\centering

  \caption{Summary of the properties of a particular model BL (where the magnetic field penetrates into the core) at different stages of evolution. We show the maximum crustal field strengths and their respective locations, as well as the ellipticity and magnetic dipole moment. The locations of the maxima are given in dimensionless spherical coordinates $(r,\theta)$, such that $r=1$ is the stellar surface. These runs were obtained for a star with mass $1.4M_{\odot}$. The stellar radius is taken to be $11.6$ km, the crust core interface is located at $10.8$ km, and the bottom of the envelope, up to which equation \eqref{eq:Hallind} has been integrated, is at $11.53$ km. The dimensionless coordinate $r$ is normalised so that at a radius of $11.53$ km we have $r=1$. Therefore the crust-core boundary is located at $r= 0.94$.}
  \begin{tabular}{lcccccc}
  \hline
Time & $|\boldsymbol{B}^{\text{max}}_{\text{pol}}|$  & Crustal location  & $|\boldsymbol{B}^{\text{max}}_{\text{tor}}|$  & Crustal location  & Ellipticity & Dipole moment \\
(yr) & ($10^{14}$ G) & & ($10^{15}$ G) & & & ($10^{30}$ G cm$^3$)\\
\hline
$0$ & $  0.12 $ & $(0.94,3.14)$ & $ 16 $ & $(0.94,1.58)$ & $\lesssim 10^{-11}$ & $7.66 $  \\
$10^{3}$   & $ 0.36 $ & $(0.97,1.35)$ & $ 16 $ & $(0.94,1.54)$ & $-3.22 \times 10^{-8}$ & $7.68 $  \\
$10^{4}$   & $ 2.6 $ & $(0.97,0.95)$ & $ 11 $ & $(0.94,1.42)$ & $-2.94 \times 10^{-7}$ & $8.09 $  \\
$10^{5}$  & $ 5.7$ & $(0.95,0.51)$ & $ 4.1 $ & $(0.94,1.47)$ & $-2.58 \times 10^{-6}$ & $8.81 $  \\
$10^{6}$  & $ 7.4 $ & $(0.94,0.46)$ & $ 3.4 $ & $(0.94,1.47)$ & $-2.50 \times 10^{-6}$ & $9.06 $  \\

\hline
\end{tabular}
\label{tab:table2}
\end{table*}

\begin{table*}
\centering

  \caption{Summary of the initial and final ($t=10^6$ yr) states of a selection of AL models (where the magnetic field is confined to the crust). We show the maximum field strengths and their respective locations, as well as the deformations (ellipticity) and dipole moments. The locations of the maxima are given in dimensionless spherical coordinates $(r,\theta)$, such that $r=1$ is the stellar surface.  The star radius is taken at $11.6$ km, the crust core interface is located at $10.8$ km, and the bottom of the envelope, up to which equation \eqref{eq:Hallind} has been integrated is at $11.53$ km. We normalize the radii to $11.53$ km $=1$. Therefore the crust-core boundary is at $0.94$.}
  \begin{tabular}{llcccccc}
  \hline
Time & Model & $|\boldsymbol{B}^{\text{max}}_{\text{pol}}|$  & Crustal location  & $|\boldsymbol{B}^{\text{max}}_{\text{tor}}|$ & Crustal location  & Ellipticity & Dipole moment\\
(yr) &  & ($10^{14}$ G) & & ($10^{15}$ G) & & & ($10^{30}$ G cm$^3$)\\
\hline
0 & AL (I) & $  5.0 $ & $(0.94,1.59)$ & $  3.0 $ & $(0.97,1.56)$ & $-4.72 \times 10^{-8}$ & $38.3 $  \\
   & AL (II) & $  5.0 $ & $(0.94,1.59)$ & $  1.0 $ & $(0.97,1.59)$ & $-1.15 \times 10^{-8}$ & $38.3 $  \\
   & AL (III) & $0.5$ & $(0.94, 1.59)$ & $  3.0 $ & $(0.97,1.56)$ & $-4.02 \times 10^{-8}$ & $3.83 $  \\
   & AL (IV) & $  0.5$ & $(0.94,1.59)$ & $  1.0 $ & $(0.97,1.59)$ & $-4.52 \times 10^{-9}$ & $3.83 $ \\
\hline
$10^6$ & AL (I) & $  2.0 $ & $(0.94,0.53)$ & $ 0.2$ & $(0.94,0.67)$ & $-2.73 \times 10^{-10}$ & $0.94 $   \\
   & AL (II) & $  2.5 $ & $(0.94,1.06)$ & $ 0.13 $ & $(0.94,1.20)$ & $-3.00 \times 10^{-9}$ & $1.61 $  \\
   & AL (III) & $2.2 $ & $(0.94,0.07)$ & $  0.32 $ & $(0.94,0.14)$ & $1.03 \times 10^{-9}$ & $0.14$  \\
   & AL (IV) & $1.8 $ & $(0.94,0.14)$ &  $  0.2$ & $(0.94,0.25)$ & $5.39 \times 10^{-10}$ & $0.33$  \\

\hline
\end{tabular}
\label{tab:table3}
\end{table*}

\begin{table*}
\centering

  \caption{Summary of the initial and final ($t=10^6$ yr) states of a selection of BL models (where the magnetic field penetrates into the core). We show the maximum field strengths and their respective locations, as well as the deformations (ellipticity) and dipole moments. The locations of the maxima are given in dimensionless spherical coordinates $(r,\theta)$, such that $r=1$ is the stellar surface.  The star radius is taken at $11.6$ km, the crust core interface is located at $10.8$ km, and the bottom of the envelope, up to which equation \eqref{eq:Hallind} has been integrated is at $11.53$ km. We normalize the radii to $11.53$ km $=1$. Therefore the crust-core boundary is at $0.94$.}
  \begin{tabular}{llcccccc}
  \hline
Time & Model & $|\boldsymbol{B}^{\text{max}}_{\text{pol}}|$  & Crustal location  & $|\boldsymbol{B}^{\text{max}}_{\text{tor}}|$ & Crustal location  & Ellipticity & Dipole moment\\
(yr) &  & ($10^{14}$ G) & & ($10^{15}$ G) & & & ($10^{30}$ G cm$^3$)\\
\hline
0 & BL (I) & $  0.6$ & $(0.94,3.14)$ & $ 16 $ & $(0.94,1.59)$ & $\lesssim 10^{-11}$ & $38.3 $  \\
   & BL (II) & $0.6$ & $(0.94,3.14)$ & $ 1.6 $ & $(0.94,1.59)$ & $\lesssim 10^{-11}$ & $38.3 $  \\
   & BL (III) & $6.0 \times 10^{-2}$ & $(0.94,3.14)$ & $ 16 $ & $(0.94,1.59)$ & $\lesssim 10^{-11}$ & $3.83 $   \\
   & BL (IV) & $6.0 \times 10^{-2}$ & $(0.94,3.14)$ & $ 1.6 $ & $(0.94,1.59)$ & $\lesssim 10^{-11}$ & $3.83 $   \\
\hline
$10^6$ & BL (I) & $  9.7 $ & $(0.94,0.74)$ & $ 3.5 $ & $(0.94,1.48)$ & $-2.37 \times 10^{-6}$ & $40.9 $  \\
   & BL (II) & $  1.5 $ & $(0.94,1.13)$ & $ 0.15 $ & $(0.94,1.45)$ & $-2.84 \times 10^{-8}$ & $37.5 $  \\
   & BL (III) & $  4.1 $ & $(0.94,0.21)$ & $ 3.5 $ & $(0.94,1.48)$ & $-2.41 \times 10^{-6}$ & $4.57 $  \\
   & BL (IV) & $  0.55 $ & $(0.94,0.78)$ & $ 0.18 $ & $(0.94,0.92)$ & $-2.40 \times 10^{-8}$ & $4.09 $  \\

\hline
\end{tabular}
\label{tab:table4}
\end{table*}

\subsection{Energy redistributions}
In this section, we calculate the energies of the multipole components of the analytic magnetic fields of the AL and BL models. We do this for two reasons: (1) to show that the initial dipole field component becomes less dominant after undergoing Hall drift and (2) to check that the system does not gain energy, which is unphysical since we do not have any external sources of energy. Given an analytic reconstruction of the magnetic field, we can calculate the energy associated with each $\ell$-mode. Defining the magnetic energy in the usual way,
\begin{equation}
E = R_{\star}^3 \int_{V}\frac {|\boldsymbol{B}|^2}  {8\pi} d^{3} x,
\end{equation}
we can decompose $E$ as a sum over each multipole, i.e. $E = \sum_{\ell} E_{\ell}$, where $E_{\ell}$ is the energy associated with
\begin{equation}
\boldsymbol{B}_{\ell} = \boldsymbol{B}|_{\kappa_{\ell} \neq 0, \kappa_{j \neq \ell} = 0}.
\end{equation}

In Figure 5, we plot the distribution of energies for the high-resolution AL model (same data as in Table 3) after a few stages of evolution. We see that initially the dipole component is dominant, as expected. After $t=10^{4}$ yr, the energy distributed between the dipole mode and a few of the higher order modes ($\ell = 46, \ldots, 51$) become comparable, and after $t=10^{6}$ yr, these higher order multipoles begin to dominate. This is reflected in the dipole moments presented in Tables 1 and 3, where we see that the dipole moment is reduced by two orders of magnitude. Nevertheless, the dominant higher order multipoles do not induce a significant ellipticity ($|\epsilon| \sim 10^{-9}$). Note that the total magnetic energy decreases significantly between $t=10^4$ yr and $t=10^6$ yr, from $\sim 10^{46}$ erg to $\sim 10^{44}$ erg. This fast dissipation feature is characteristic of magnetic fields that are completely confined to the crust \citep{KK12,GV14}. Due to the relatively low mass density in the crust, the typical length-scale of the magnetic field is reduced dramatically by Hall drift, which in turn enhances Ohmic dissipation. Moreover, Joule heating becomes very efficient because the magnetic field length-scales become smaller, thus amplifying the effects of the finite and temperature-dependent electrical conductivity \citep{PG07,VRPPAM13}. The bulk of the magnetic energy is dissipated as heat, increasing the thermal luminosity of the star \citep{VRPPAM13}. Some of the magnetic energy is transformed into elastic stresses which are eventually released in bursts \citep{PP11,PPX11}. Dissipation in the magnetosphere, which is twisted just above the magnetic spots, may also play a role \citep{GCFMS13}. It is important to note that energy is indeed conserved in all these mechanisms. However, the avenues of magnetic energy dissipation and their relative importance are beyond the scope of this work.


In Figure 6, we plot the distribution of energies for the high-resolution BL model (same data as in Table 2) after a few stages of evolution. We see that initially the dipole component is dominant, as expected, while over time the energy is distributed mostly between the dipole mode and a few of the higher order modes ($\ell = 48, 51, 52$ and $53$). The total energy decreases over time, indicating that some of the dipole energy stored in the toroidal reservoir is being redistributed to poloidal high-$\ell$ modes, though not as dramatically as in the AL case. Even after $t=10^{6}$ yr, the dipole is still the energetically dominant mode.

Figures 5 and 6 confirm that Hall drift tends to build up the smaller scale, higher order field components over time, at the expense of the large-scale dipole component (as expected) and that the total magnetic fields do not grow unphysically. 

Note that any magnetic field constructed on a numerical grid, such as those given as the output of the Alicante code, will have an associated multipolar resolution, i.e. there will exist an $\ell_{\text{max}}$, dependent on the grid resolution, beyond which the magnetic field cannot be resolved. This is effectively a consequence of the Shannon-Nyquist theorem \citep{nyq28}. While the nature of the staggered, logarithmic grids used in the Alicante code makes a grid-to-spectral comparison difficult, we find that adding or subtracting up to 10 multipole orders ($N \rightarrow N \pm 10$) does not quantitatively affect the energy distributions by more than a few percent. Though not shown here, we also find no evidence that the maximum spectral resolution has been exceeded in the least squares fitting algorithm (detailed in Sec 3.2) in the `lower'-resolution runs. We can conclude then that our analysis is not restricted by such grid-to-spectral considerations, though one should be wary if trying to resolve large ($N \gtrsim 100$) multipole orders. 
\begin{figure*}
\centering
\begin{subfigure}[b]{0.33\textwidth}
\includegraphics[width=\textwidth]{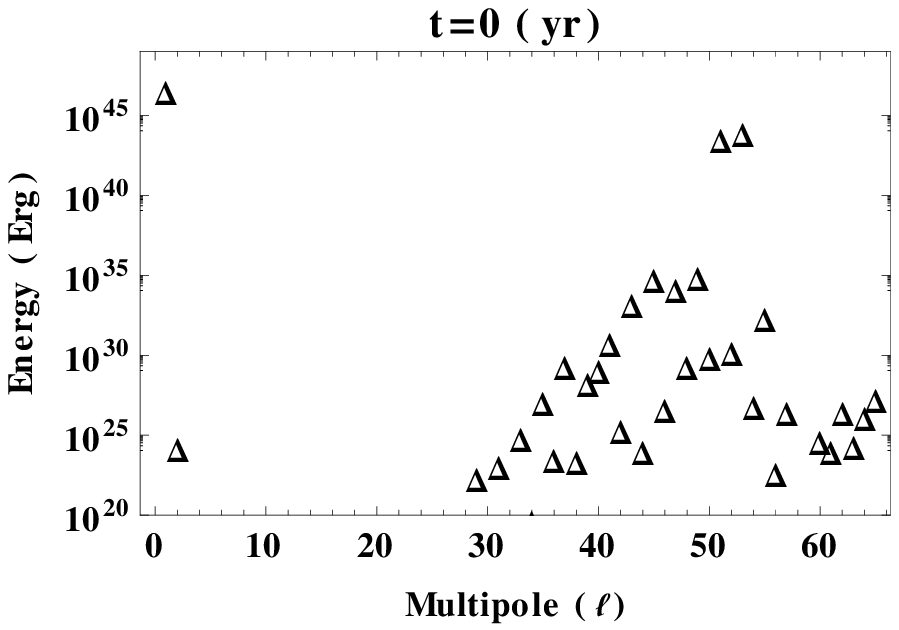}
\end{subfigure}
\begin{subfigure}[b]{0.33\textwidth}
\includegraphics[width=\textwidth]{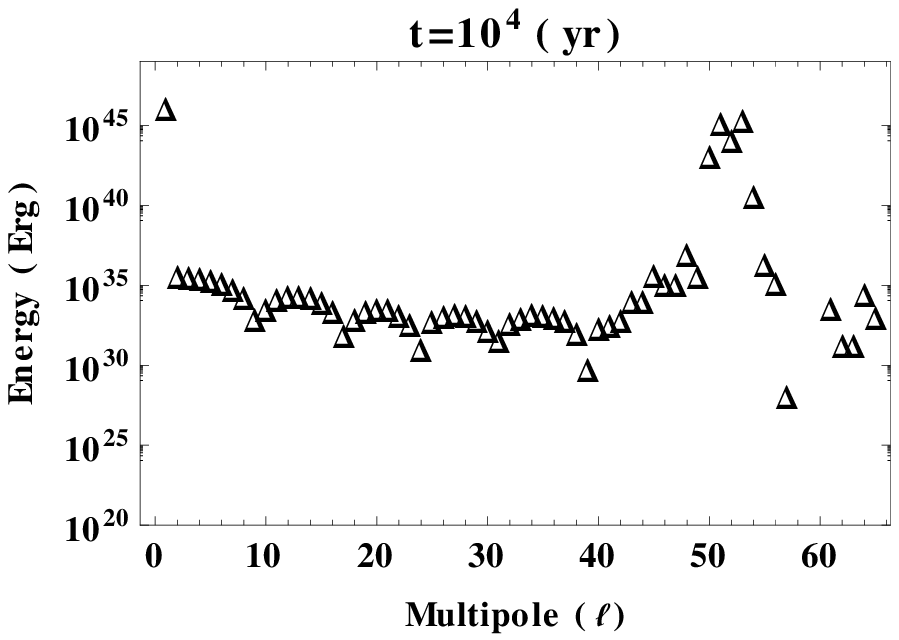}
\end{subfigure}
\begin{subfigure}[b]{0.33\textwidth}
\includegraphics[width=\textwidth]{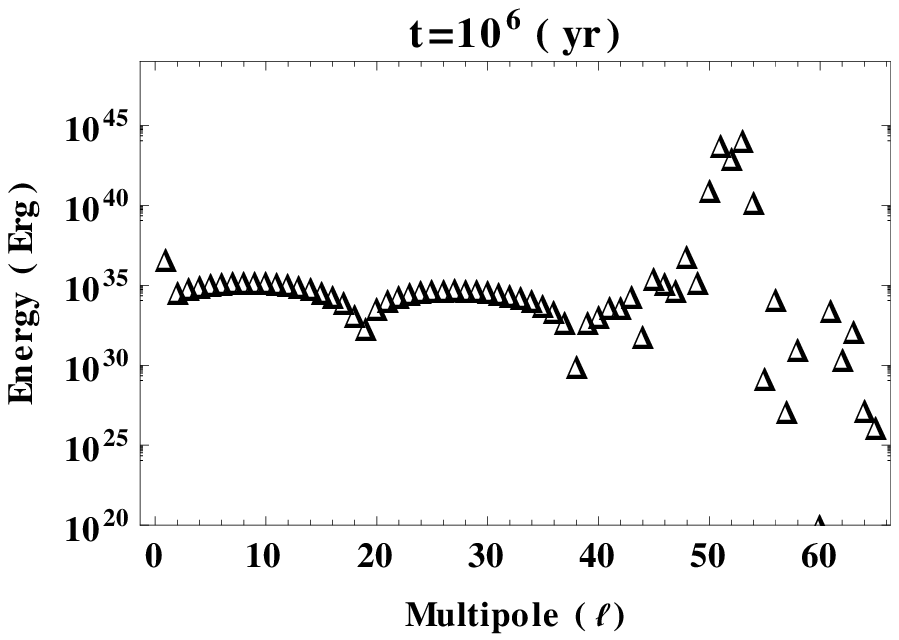}
\end{subfigure}
\caption{Energy stored in individual $\ell$-modes for model AL at times $t=0$ (left), $t=10^{4}$ (yr) (middle), and $t=10^{6}$ (yr) (right). The \emph{total} energies stored in each case are $E= 2.79 \times 10^{46}$ erg, $E= 1.63 \times 10^{46}$ erg, and $E= 1.92 \times 10^{44}$ erg, respectively. }
\label{fig:AL_energies}
\end{figure*}

\begin{figure*}
\centering
\begin{subfigure}[b]{0.33\textwidth}
\includegraphics[width=\textwidth]{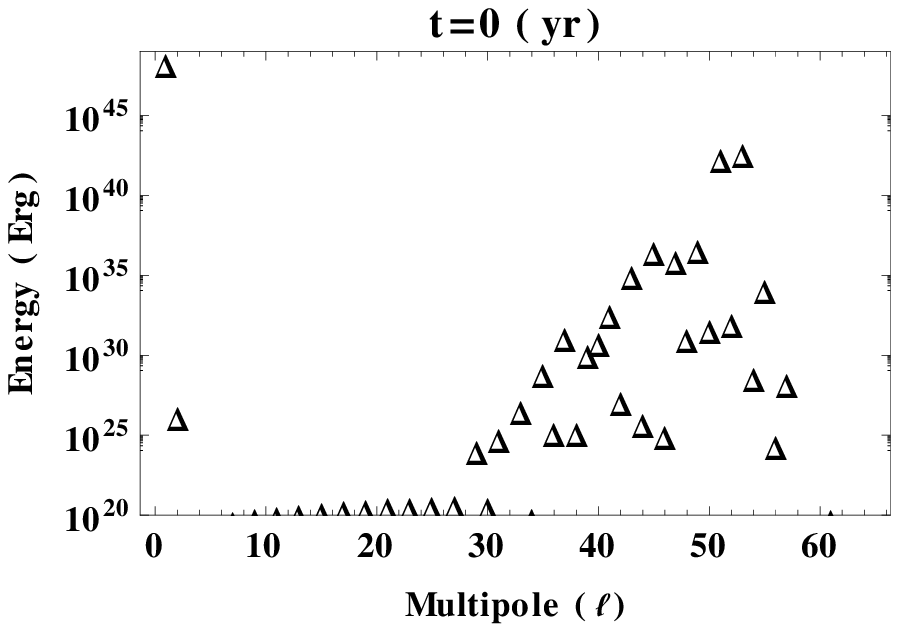}
\end{subfigure}
\begin{subfigure}[b]{0.33\textwidth}
\includegraphics[width=\textwidth]{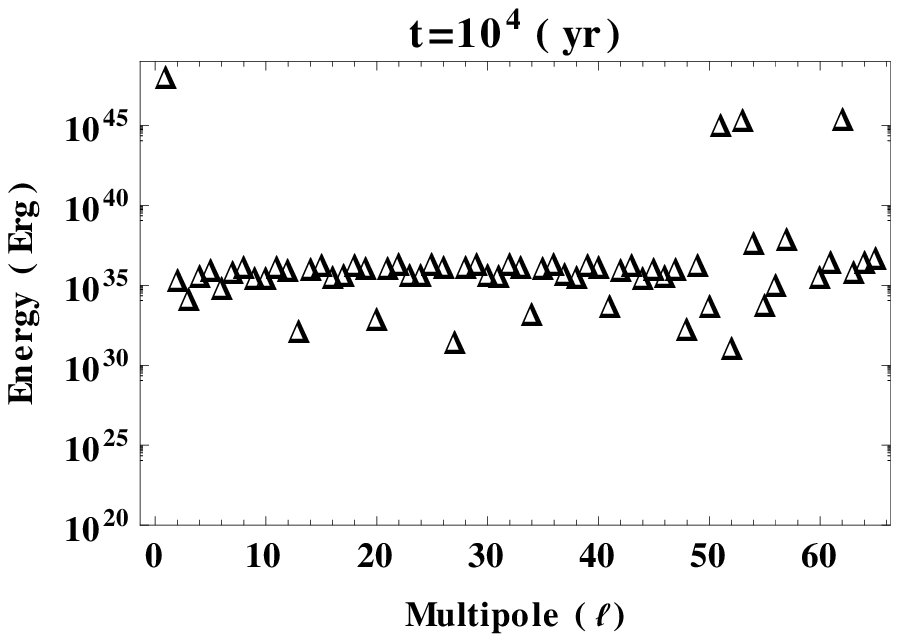}
\end{subfigure}
\begin{subfigure}[b]{0.33\textwidth}
\includegraphics[width=\textwidth]{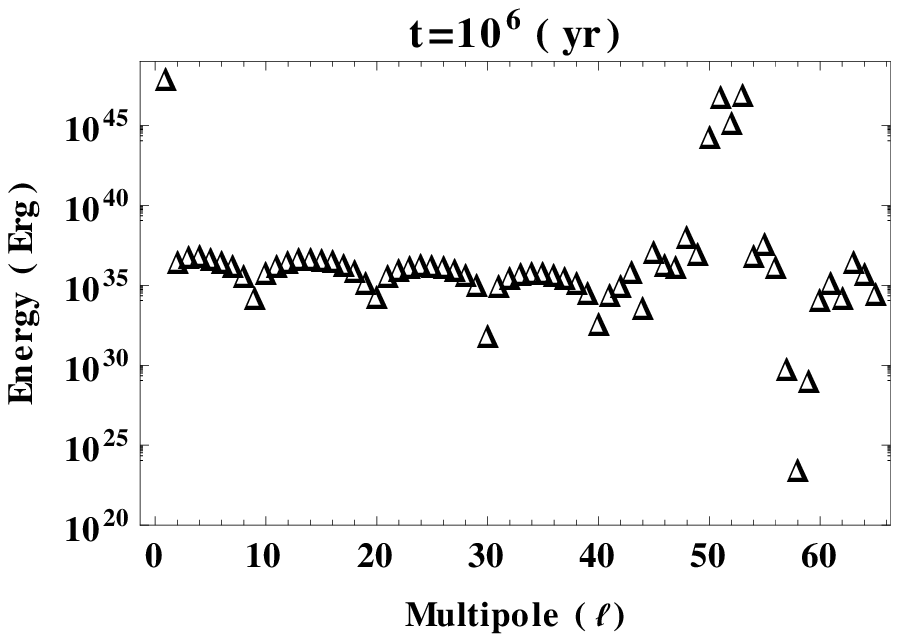}
\end{subfigure}
\caption{Energy stored in individual $\ell$-modes for model BL at times $t=0$ (left), $t=10^{4}$ (yr) (middle), and $t=10^{6}$ (yr) (right). The \emph{total} energies stored in each case are $E= 1.44 \times 10^{48}$ erg, $E= 1.30 \times 10^{48}$ erg, and $E= 1.15 \times 10^{48}$ erg, respectively.}
\label{fig:BL_energies}
\end{figure*}

\section{Gravitational radiation}

Gravitational waves are generated by a rotating, biaxial (i.e., $\epsilon \neq 0$) star when its `wobble angle' (the angle between its total angular momentum vector and symmetry axis) is non zero. The most general expression for the GW signal [e.g., given by \citet{JKS98}] depends on $\epsilon$, wobble angle, and the `line-of-sight angle' (the angle between the angular momentum vector and the line of sight to the observer). The signal is strongest when the wobble angle is $\pi/2$ and the line-of-sight angle is zero. If we assume that the orientation is optimal, we can write the dimensionless gravitational wave strain $h_0$ as \citep{AETAL10}

\begin{equation} \label{eq:h0}
h_0 = \textrm{ }4.2\times 10^{-26} \left(\frac{\nu}{100\textrm{ Hz}}\right)^2 \left(\frac{I_0}{10^{45}\textrm{ g cm}^2}\right) \left(\frac{|\epsilon|}{10^{-6}}\right) \left(\frac{d}{1\textrm{ kpc}}\right)^{-1} \\
\end{equation}
where $\nu$ is the spin frequency and $d$ is the distance to the star. If we assume that all the spin-down luminosity of a pulsar comes from gravitational wave radiation, we can set the canonical `spin-down limit' on wave strain $h_0^{\textrm{sd}}$ \citep{aasi14}

\be h_0^{\textrm{sd}} = 8.1 \times 10^{-19} \left(\frac{I_0}{10^{45}\textrm{ g cm}^2}\right)^{1/2} \left(\frac{d}{1\textrm{ kpc}}\right)^{-1} \left(\frac{|\dot\nu / \text{Hz} \text{ s}^{-1}|}{\nu / \text{Hz}}\right)^{1/2},\ee
where $\dot\nu$ is the frequency derivative of the pulsar.

\citet{aasi14} presented the results from the latest science runs of initial-generation GW detectors LIGO (Laser Interferometric Gravitational-wave Observatory) and Virgo. No evidence of GW were detected. However, they highlighted 7 pulsars whose observed upper limits $h_0$ are within a factor of 4 of their $h_0^\textrm{sd}$. In particular, for the Crab and Vela pulsars, \citet{aasi14} found $h_0<h_0^\textrm{sd}$. What do the AL and BL models predict for these 7 pulsars and how do they compare to observations?

Assuming the star to act as an orthogonal rotator and bounding the electromagnetic braking energy by the rotational kinetic energy loss, we can estimate the surface magnetic field for a pulsar (e.g. \cite{ST83})
\begin{equation} \label{eq:charb}
|\boldsymbol{B}_{\text{s}}| > \Bigg| \frac {3 c^3 I_{0}} {8 \pi^2 R^{6}}  \frac {\dot{\nu}} {\nu^3} \Bigg|^{1/2}.
\end{equation}
Also, the characteristic age formula for a pulsar with braking index $n=3$ is given by,
\begin{equation} \label{eq:chart}
\tau_{\text{c}} = |\frac { \nu} {2 \dot{\nu}}|.
\end{equation}
Using the above estimates in comparison with our time evolved models with various magnetic field strengths, we can compare our predicted values for $\epsilon$ for both the AL and BL models for the 7 high-interest pulsars of \citet{aasi14}. We present the results in Table 5.

As expected, the BL model generally predicts higher $|\epsilon|$ than the AL model, by about two orders of magnitude. For the two youngest pulsars (J0537$-$6910 and J1833$-$1034) and the two pulsars with the lowest $\bsb_s$ (J1913$+$1011 and J1952$+$3252), the AL and BL models predict similar $\epsilon$. Note that equation \eqref{eq:charb} presumes a dipolar magnetic field, while in our models we have multipolar fields with strong non-dipole components (cf. Figures 5 and 6). The actual $|\boldsymbol{B}^{\text{max}}_{\text{pol}}|$ values of these $7$ objects may be higher than the values presented in Table \ref{tab:table5}. Table 5 shows that our results are not ruled out by observations.

As equation (\ref{eq:h0}) shows, $h_0$ is directly proportional to $\epsilon$. This suggests that magnetars, with $\epsilon\sim 10^{-6}$ \citep{MMRA11}, should be the best potential sources of GW. However, magnetars have $\nu\sim 0.1$ Hz,\footnote{For an up-to-date catalogue of magnetars, see http://www.physics.mcgill.ca/~pulsar/magnetar/main.html \citep{OK14}.} implying $h_0\lesssim 10^{-28}$. Furthermore, seismic, thermal, and quantum noises reduce the sensitivity of GW detectors at low frequencies, $\nu\lesssim 10$ Hz \citep{AETAL10,HETAL11}. Hence, traditionally Galactic magnetars are not considered to be prime targets, although newborn magnetars are more promising \citep{TCQ04,SETAL05,DSS09}.

However, we see in Table \ref{tab:table4}, for example, that Hall drift changes the situation. It is possible for a neutron star with a dipolar field structure and with initial surface field strength of $\sim 10^{13}$ G, to develop a strong, localised magnetic spot and $\epsilon\sim 10^{-6}$ after undergoing Hall-drift induced magneto-thermal evolution for $t\sim 10^{5}$ yr. Interestingly, this means that younger magnetars are better candidates than older magnetars, but older pulsars (where Hall drift has had more time to operate and evolve the magnetic fields) are better candidates than younger pulsars.

\begin{table*}
\centering

  \caption{Comparison of our models with observational limits of 7 selected pulsars, for which $h_0\lesssim 4 h_0^\textrm{sd}$ \citep{aasi14}. $|\boldsymbol{B}_{s}|$ and $\tau_{c}$ were computed from data given in {\protect\cite{aasi14}.}  The fourth column shows the observational upper limits on $\epsilon$ from LIGO and Virgo \citep{aasi14}, the fifth column shows $\epsilon$ as predicted by the AL model, and the sixth column shows $\epsilon$ as predicted by the BL model.}
  \begin{tabular}{llcccc}
  \hline
Pulsar & $|\boldsymbol{B}_{\text{s}}|$ & $\tau_{\text{c}}$  & Obs. limit  $|\epsilon|$ & AL-predicted $|\epsilon|$ & BL-predicted $|\epsilon|$  \\
 & ($10^{12}$ G) & (yr) &   & &  \\
\hline
J$0534$+$2200$ (Crab) & $1.84$ & $1.3 \times 10^{4}$ & $8.6 \times 10^{-5}$ & $9.6 \times 10^{-10}$ &  $2.9 \times 10^{-7}$   \\
J$0537$-$6910$        & $1.42$  & $4.9 \times 10^{3}$ & $1.2 \times 10^{-4}$ & $5.8 \times 10^{-8}$ &  $3.2 \times 10^{-8} $     \\
J$1813$-$1246$        & $1.43$  & $4.3 \times 10^{4}$ & $3.5 \times 10^{-4}$ & $9.6 \times 10^{-10}$  & $2.9 \times 10^{-7}$   \\
J$1833$-$1034$        & $5.52$  & $4.8 \times 10^{3}$ & $5.7 \times 10^{-3}$ & $5.8 \times 10^{-8}$  & $3.2 \times 10^{-8} $   \\
J$1913$+$1011$        & $0.54$  & $1.7 \times 10^{5}$ & $2.2 \times 10^{-4}$ & $1.3 \times 10^{-8} $ & $ 2.6 \times 10^{-6} $   \\
J$0835$-$4510$ (Vela) & $5.26$  & $1.1 \times 10^{4}$ & $6.0 \times 10^{-4}$ & $9.6 \times 10^{-10}$  & $2.9 \times 10^{-7}$   \\
J$1952$+$3252$        & $0.74$  & $1.1 \times 10^{5}$ & $3.0 \times 10^{-4}$ & $1.3 \times 10^{-8} $ & $ 2.6 \times 10^{-6} $   \\

\hline
\end{tabular}
\label{tab:table5}
\end{table*}

\section{Discussion and conclusions.}

In this paper, we show that polar, spot-like magnetic field structures necessary for the functioning of radio pulsars, which are created naturally via the Hall drift for a range of initial conditions, can give rise to $|\epsilon| \gtrsim 10^{-6}$. A typical evolution involves the toroidal field redistributing itself to lead to the creation of high-order multipolar structures in the poloidal field, with magnetic energies comparable to the dipole component. For example, as shown in Table \ref{tab:table4}, the BL model with initial maximum poloidal field strength of $10^{13}$ G, well below magnetar field strength $\sim 10^{15}$ G, can develop a magnetic spot with maximum poloidal field strength of $8 \times 10^{14}$ G at the crust-core interface and at a meridional angle of about $46^\circ$  after $10^6$ yr. The density perturbation caused by this field structure is enough to deform the star into a prolate shape with $\epsilon \sim 10^{-6}$.

The crustal toroidal field strength actually decreases during the above process. In other words, in this particular setup, the magnetic spot takes over the role of the internal toroidal field in deforming the star [cf., for example, the results of \citet{MMRA11} for a purely dipolar, hemispherically symmetric magnetic field]. This is more clearly demonstrated by model BL (III) (see third and seventh rows of Table \ref{tab:table2}), where $|\boldsymbol{B}^{\text{max}}_{\text{pol}}|$ increases by two orders of magnitude, but the toroidal maximum decreases by $80\%$, resulting in a prolate star with $\epsilon \sim 10^{-6}$.

In contrast, while the magnetic spot also emerges in the AL model, the resulting magnetic field structure is only enough to deform the star into $|\epsilon| \lesssim 10^{-9}$. In fact, the magnetic spot in the AL model tends to deform the star less than the initial field configuration (i.e., $|\epsilon|$ decreases over time), as shown in Tables \ref{tab:table1} and \ref{tab:table3}. The major reason for the discrepancy is that, without a magnetic field in the core, continuity demands that the field tend to zero on the crust-core interface. Instead of creating strong gradients near the boundary, the effect is spread across the entire crust, resulting in a more uniform field, which produces a smaller deformation.

For favourable orientations, those radio pulsars whose magnetic field is not only confined to the crust but penetrates continuously the whole star may thus be observable in GW. This means, that future GW detections of radio pulsars will give a valuable hint on the internal magnetic field structure of neutron stars, which cannot be provided by electromagnetic observations. Taking the caveats into account (see Sec. 2.1) for the magneto-thermal evolution, the study presented here can provide only an indication that radio pulsars are potential sources of observable GW. Since this signal is persistent and the locations and ephemerides of radio pulsars are well known, such a signal, together with the analysis presented here, will open another window into the internal field structure. The results summarized in Tables \ref{tab:table2} and \ref{tab:table4} suggest that, even with the magnetic spots, only pulsars with ages $\gtrsim 10^5$ yr may have $|\epsilon| \sim 10^{-6}$. An exhaustive survey covering all possible initial states and parameters is beyond the scope of this paper, whose primary aim is to show the potential effects of Hall drift on magnetic field configurations and how they can be detected. As seen in Table 5, our results, applied to 7 pulsars highlighted by \cite{aasi14}, are not ruled out by current observations.

Our analysis is limited by two assumptions: (i) magnetic field axisymmetry and (ii) the simplified evolution of the core field. An extension to fully three-dimensional modelling will modify the magnetic spots (which are then truly spots rather than annuli around the pole). In this paper, the core field evolution, important for the BL model, is simply modelled by slow diffusion; the nature of the core is largely unknown. Magnetic flux expulsion from the core into the crust may continuously replenish magnetic energy into the spot region. In future, the study presented here should be extended by modelling the magnetic field evolution in three dimensions and by using a more realistic description of the magnetic energy transfer from the core into the crust. In this paper we have considered continuous GW emissions, though it may also be interesting to consider the superimposed stochastic GW contributions from all pulsars undergoing Hall drift, following calculations along the lines of those presented in \citet{LBM13}. If locally strong magnetic fields are hidden beneath the surfaces of radio pulsars, the contribution from the magnetic field to stochastic GW calculations may be underestimated.

\section*{Acknowledgments}
We thank Andrew Melatos for many valuable discussions. We thank the anonymous referee for his/her diligence in providing carefully considered feedback, which improved the quality of the manuscript. This work was supported in part by an Australian Research Council Discovery Project Grant (DP110103347), an Australian Postgraduate Award and by the grants DEC-2012/05/B/ST9/03924 of the Polish National Science Center. The generosity of D. Vigan\`{o}, J.A. Pons, and J.A. Miralles to use the Alicante magneto-thermal evolution code is gratefully acknowledged.

\bibliography{pulsars}

\appendix

\section{Newtonian and Cowling approximations}
Within the scope of the methods presented in the main text, there are three primary sources of internal error in estimating the ellipticity: (i) errors obtained during the fitting procedure (quantified in Sec. 3.3), (ii) errors introduced from taking the Cowling approximation, and (iii) errors from matching a Newtonian equilibrium to a general relativistic output that the Alicante magneto-thermal code generates. The presence of the gravitational redshift factor $e^{\nu}$ appearing in the induction \eqref{eq:Hallind} and thermal evolution \eqref{eq:Tevol} equations couples directly to the magnetic field \eqref{eq:magfield22}.

We expect the errors in $\epsilon$ associated with the Cowling approximation to be less than an order of magnitude \citep{Y13}. \citet{Y13} found that taking the Cowling approximation, when the background density profile is polytropic, only alters $\epsilon$ by a factor of $\lesssim 2$, for a dipolar mixed poloidal-toroidal magnetic field configuration. While we have higher order multipoles present in the fitting procedure (Sec. 3), the dipole field typically continues to be the dominant contributor to the magnetic energy, particularly in model BL where the dipole moment evolves only by a few per cent over $10^{6}$ years.
Reverting back to SI units to demonstrate the appearance of $\mu_{0}$ explicitly, we find that introducing the perturbed gravitational potential $\delta \Phi$ results in the following modification of equation (7) \citep{Y13}

\begin{equation}
\label{eq:modforce}
\frac{\partial\delta\rho}{\partial\theta} + \frac {\partial \delta \Phi} {\partial \theta} \frac {d \rho} { d r} \left(\frac {d \Phi} {d r}\right)^{-1} = -\frac{r}{\mu_{0} \rstar} \left( \frac {d \Phi} { dr} \right)^{-1} \{\nabla\times[(\nabla\times{\bf{B}}\times{\bf{B}})]\}_\phi
\end{equation}
The term $\delta \Phi$ has two separate components here, the first comes from the Eulerian perturbation ($\rho \rightarrow \rho + \delta \rho$), and the other from the Einstein factor $e^{\nu}$ in equations \eqref{eq:Hallind} and \eqref{eq:Tevol}. In essence, we have two expansions, the perturbative expansion in the fluid elements and the post-Newtonian expansion in the gravitational potential. To express this, we write equation \eqref{eq:modforce} as
\begin{equation}
\label{eq:modforce2}
\begin{aligned}
&\frac{\partial\delta\rho^{\text{Euler}}}{\partial\theta} + \frac{\partial\delta\rho^{\text{PN}}}{\partial\theta}  +  \left[ \frac {\partial \delta \Phi^{\text{Euler}}} {\partial \theta} + \frac {\partial \delta \Phi^{\text{PN}}} {\partial \theta} \right] \frac {d \rho} { d r} \left(\frac {d \Phi} {d r}\right)^{-1} \\
&=-\frac{r}{\mu_{0} \rstar} \left( \frac {d \Phi} { dr} \right)^{-1} \{\nabla\times[(\nabla\times{\bf{B}}\times{\bf{B}})]\}_\phi ,
\end{aligned}
\end{equation}
where $\delta \Phi^{\text{Euler}}$ is given through the perturbed Poisson equation $\nabla^2 \delta \Phi^{\text{Euler}} = 4 \pi G \delta \rho^{\text{Euler}}$, with $\delta \rho^{\text{Euler}}$ satisfying equation \eqref{eq:modforce}.

In light of Yoshida's (2013) results, we expect $\delta \Phi^{\text{Euler}}$ to be small, but it remains to quantify the magnitude of the Einstein contribution to the ellipticity. One can estimate the leading order contribution from the gravitational redshift factor $e^{\nu}$ to the perturbed density profile $\delta \rho^{\text{PN}}$ by writing the $tt$-component of the metric as follows \citep{W84}
\begin{equation}
\label{eq:postnew}
g_{tt} = e^{\nu} = 1 - \frac {2 \Phi} {c^2} - \frac {2 \delta \Phi^{\text{PN}}} { c^2}.
\end{equation}
Taking the Eulerian perturbation $\delta \rho^\text{Euler}$ to satisfy equation (A1), we find that the post-Newtonian density can be estimated by substituting equation (A3) into (A2):

\begin{equation}
\label{eq:relns}
\delta \rho^{\text{PN}} =  - \frac {c^2} {2} \left( 1 - e^{\nu} - \frac {2 \Phi} {c^2} \right)  \frac {d \rho} { dr } \left(\frac {d \Phi} {d r}\right)^{-1} .
\end{equation}
Given the values of $e^{\nu}$ from the Alicante code and a background gravitational potential $\Phi$, one can estimate the Einstein contribution to the deformation \eqref{ellipticity}. Note that we have $\delta \rho^{\text{PN}} \propto d \rho / dr$.

For both the AL and BL models, we assume an $n=1$ polytrope profile \eqref{eq:polytrope} for the background density $\rho$. This choice is made primarily because then we find that
\begin{equation}
c^2 \frac {d \rho} { d r} \left( 1 - e^{\nu} - \frac {2 \Phi} {c^2} \right) \sim 10^{-16} \rho\Phi.
\end{equation}
As a consequence of the above estimate and equations \eqref{eq:modforce2} and \eqref{eq:relns}, we find that the dimensionless ratio $\delta \rho^{\text{PN}} / \delta \rho^{\text{Euler}}$ reads
\begin{equation} \label{eq:aboverho}
\begin{aligned}
\delta \rho^{\text{PN}} / \delta \rho^{\text{Euler}} &\sim 10^{-5} \left( \frac {\rho} {10^{17} \text{ kg} \text{ m}^{-3}} \right) \left( \frac {\Phi} {10^{12} \text{ m} \text{ s}^{-2}} \right) \\
&\times \left( \frac {R_{\star}} {10^{4} \text{ m}} \right) \left( \frac {|\bsb|} {10^{8} \text{ T}} \right)^{-2} .
\end{aligned}
\end{equation}
This ratio is $\ll 1$ in the neutron star regime. As such, we are justified in employing the simplified Newtonian framework in Sec. 2.2 and beyond, as opposed to using the full general relativistic treatment that the Alicante code produces, as described in Sec. 2.

\end{document}